\documentclass[useAMS,usegraphicx]{mn2e}

\title[What is the origin of the soft excess in AGN?]{What is the
  origin of the soft excess in AGN?}

\author[Ma{\l}gorzata A. Sobolewska and Chris Done]{Ma{\l}gorzata
A. Sobolewska\thanks{E-mail:m.a.sobolewska@durham.ac.uk} 
and Chris Done\thanks{E-mail:chris.done@durham.ac.uk}\\
Department of Physics, University of Durham, South Road, Durham
  DH1 3LE}
\newcommand{\apj}{ApJ}
\newcommand{\apjl}{ApJL}
\newcommand{\mnras}{MNRAS}
\newcommand{\aap}{A\&A}

\newcommand{\pasj}{PASJ}

\begin{document}

\date{}

\pagerange{\pageref{firstpage}--\pageref{lastpage}} \pubyear{2006}

\maketitle

\label{firstpage}

\begin{abstract}
We investigate the nature of the soft excess below 1~keV observed in
AGN. We use the {\it XMM-Newton} data of the low redshift, optically
bright quasar, PG~1211+143, and we compare it with the Narrow Line
Seyfert 1 galaxy, 1H~0707-495, which has one of the strongest soft
excesses seen. We test various ideas for the origin of the soft X-ray
excess, including a separate spectral component (for example, low
temperature Comptonized emission), a reflection-dominated model, or a
complex absorption model. All three can give good fits to the data,
and $\chi^2$ fitting criteria are not sufficient to discriminate among
them. Instead, we favour the complex absorption model on the grounds
that it requires less extreme parameters. In particular the geometry
appears to be more physically plausible as the reflected component in
the smeared absorption model is no longer dominant, and relativistic
distortions, while still clearly present, are not tremendously larger
than expected for a disc around a Schwarzchild black hole.

\end{abstract}


\section{Introduction}

The X-ray spectra of Active Galactic Nuclei (AGN) and Quasars have
long been known to often contain a 'soft excess' component at energies
below $\sim$~1~keV (e.g. Turner \& Pounds 1988). This rises rather
smoothly above the extrapolation of the power law continuum seen in
the 2--10~keV band, and seems to connect onto the peak of the UV
accretion disc spectrum (Zheng et al. 2001; Czerny et al. 2003). This shape
is suggestive of low temperature, high optical depth Comptonization of
the inner accretion disc, in addition to the high temperature, low
optical depth Comptonization required to produce the high energy power
law (e.g. Page et al. 2004a). However, the temperature of this
additional component seems remarkably constant in high mass accretion
rate objects ($L/L_{Edd}>0.1$), despite a large range in black hole
mass and hence inferred accretion disc temperature (Czerny et al.
2003; Gierli\'nski \& Done 2004, hereafter GD04; Crummy et
al. 2006). The nature of this soft excess becomes  
even more puzzling when compared to the spectra of Galactic black hole
(GBH) binary systems. These can indeed show a low temperature, high
optical depth Comptonized component in addition to the disc and high
energy power law spectrum (in the very high state).  However, while it
is initially tempting to identify this as the GBH counterpart of the
soft excess seen in AGN (Page et al. 2004b; Murashima et al. 2005), the
parameters required are rather different, most obviously in the fact
that the temperature of this component is {\em variable} in the GBH
(Kubota, Makishima \& Ebisawa 2001; Kubota \& Done 2004).

AGN with the largest soft excesses, generally Narrow Line Seyfert 1
galaxies (NLS1; Boroson 2002), also often show a strong
deficit at $\sim 7$~keV which again has no obvious identification
(Boller et al. 2002; Fabian et al. 2002, 2004, 2005; Tanaka, Ueda \&
Boller 2005). These objects are often highly luminous, 
close to the Eddington limit, yet their underlying high energy
2--10~keV spectrum can appear to be rather hard
(e.g. GD04; Crummy et al. 2006). All Galactic black holes at
$L/L_{Edd}\sim 0.5$--1 have intrinsically soft power law Comptonized
emission (e.g. Remillard \& McClintock 2006), though occasional
super Eddington source spectra can be harder (e.g. V404 Cyg: \.Zycki,
Done \& Smith 1999). 

The constancy of temperature for the soft excess emission could
plausibly point to an origin in atomic processes. In particular, there
is an abrupt increase in opacity in partially ionized material between
$\sim 0.7$--3~keV due to OVII/OVIII and Fe L transitions which can
produce a large increase in reflected or transmitted flux below
0.7~keV, making the apparent soft X-ray excess. The corresponding rise
in reflected or transmitted flux above 2--3~keV could also
artificially harden the high energy spectrum, while the same material
might also have strong iron features, giving a simultaneous
explanation of the 7~keV deficit.

By contrast, models where the soft excess is a separate continuum
component require some other explanation for the origin of the iron
features and hard spectra, such as partial covering of the source by
cold absorbing material (Boller et al. 2002; Tanaka et al. 2004). However,
the neutral iron edge at 7.1~keV in these models is not big enough to
fit the strongest observed features unless iron is very overabundant,
at least 5--7 times the Solar value
(Tanaka et al. 2004; Gallo et al. 2004; Gallo 2006). More generally, such
models seem contrived as the iron features and hard spectra appear to
be correlated with the soft excess, pointing to a common origin for
the hard and soft spectral complexity (Boller et al. 2003; Reeves,
Porquet \& Turner 2004; Tanaka et al. 2005). This connection is 
strongly supported by spectral variability. The {\it rms} spectra from
these objects typically show more variability in the 0.7--3~keV range
than at higher or lower energies
(e.g. Gallo et al. 2004; Ponti et al. 2004, 2006). The spectrum
below 0.7~keV, which is 
dominated by the soft excess, typically has the same variability as
the spectrum above 3~keV, showing that they are almost certainly
linked to the same component. The amplified variability in the
0.7--3~keV region is compelling evidence for an atomic origin for the
spectral complexity, as it is exactly over this energy range that the
atomic features dominate (Gierli\'nski \& Done 2006), but both
reflection and absorption can match the variability spectra seen
(Ponti et al. 2006; Gierli\'nski \& Done 2006).

One obvious problem with the atomic models is that they
predict strong, sharp atomic features from the partially ionized
material. Such characteristic narrow absorption lines and sharp edges
are seen in AGN (warm absorbers) but these can be easily isolated
using high spectral resolution data (see e.g. Blustin et al. 2005 and
references therein), and a strong {\em smooth} soft excess
remains. Large velocity smearing is the only way to keep an atomic
origin of the soft excess. If at least some of the material is moving at
moderately relativistic speeds, the characteristic line/edge
features become strongly broadened into a quasi--continuum. Such speeds
naturally arise only close to the black hole, so the atomic models
predict that the soft excess is formed in regions of strong gravity,
irrespective of whether it arises from partially ionized reflection or
absorption. However, it is important to distinguish between a reflection or
absorption origin for the soft excess, as they give very different
geometries and have different implications for our understanding of
the X-ray source. In reflection, the partially ionized material is
optically thick and out of our line of sight. The accretion disc is
the obvious identification but in order to produce the strongest soft
excesses a reflection model requires that the intrinsic continuum is
strongly suppressed, perhaps by the disc fragmenting or by extreme
lightbending
(Fabian et al. 2002, 2004, 2005; Miniutti \& Fabian 2004). The
velocity structure should be Keplarian, and the large inferred
smearing implies a very small inner disc radius, and hence extreme
Kerr space-time and/or extremely centrally concentrated emissivity 
(e.g. Miniutti \& Fabian 2004).  By contrast, in an
absorption origin, the partially ionized material is optically thin,
and seen in our line of sight. This implies that it is some sort of
wind from the disc, so its velocity structure is not well defined, and
cannot easily give constraints on the space-time
(GD04; Chevallier et al. 2006; Schurch \& Done 2006).


\begin{figure*}
\centering
\includegraphics[width=8.8cm,bb=32 144 545 444,clip]{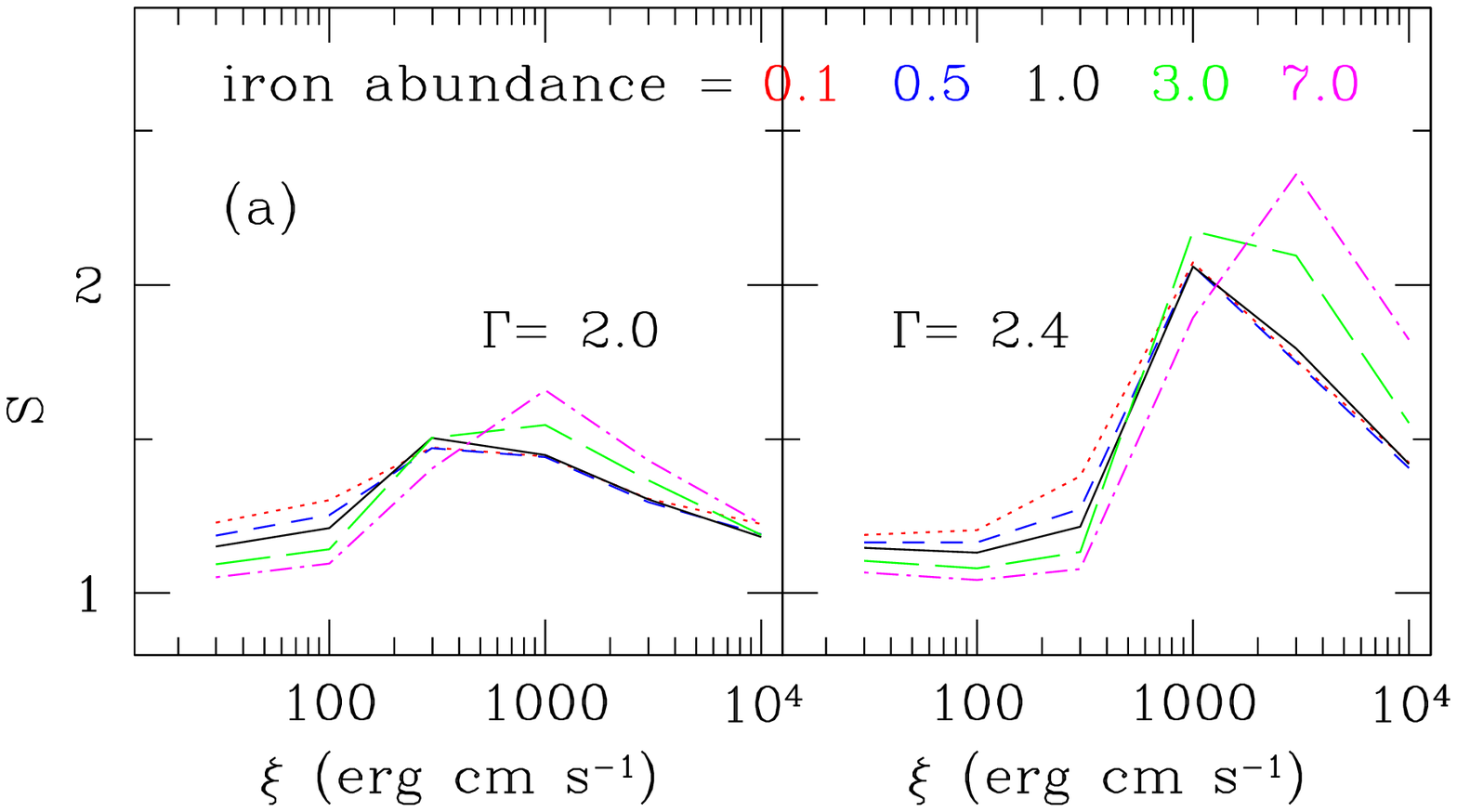}
\includegraphics[width=8.8cm,bb=32 144 545 444,clip]{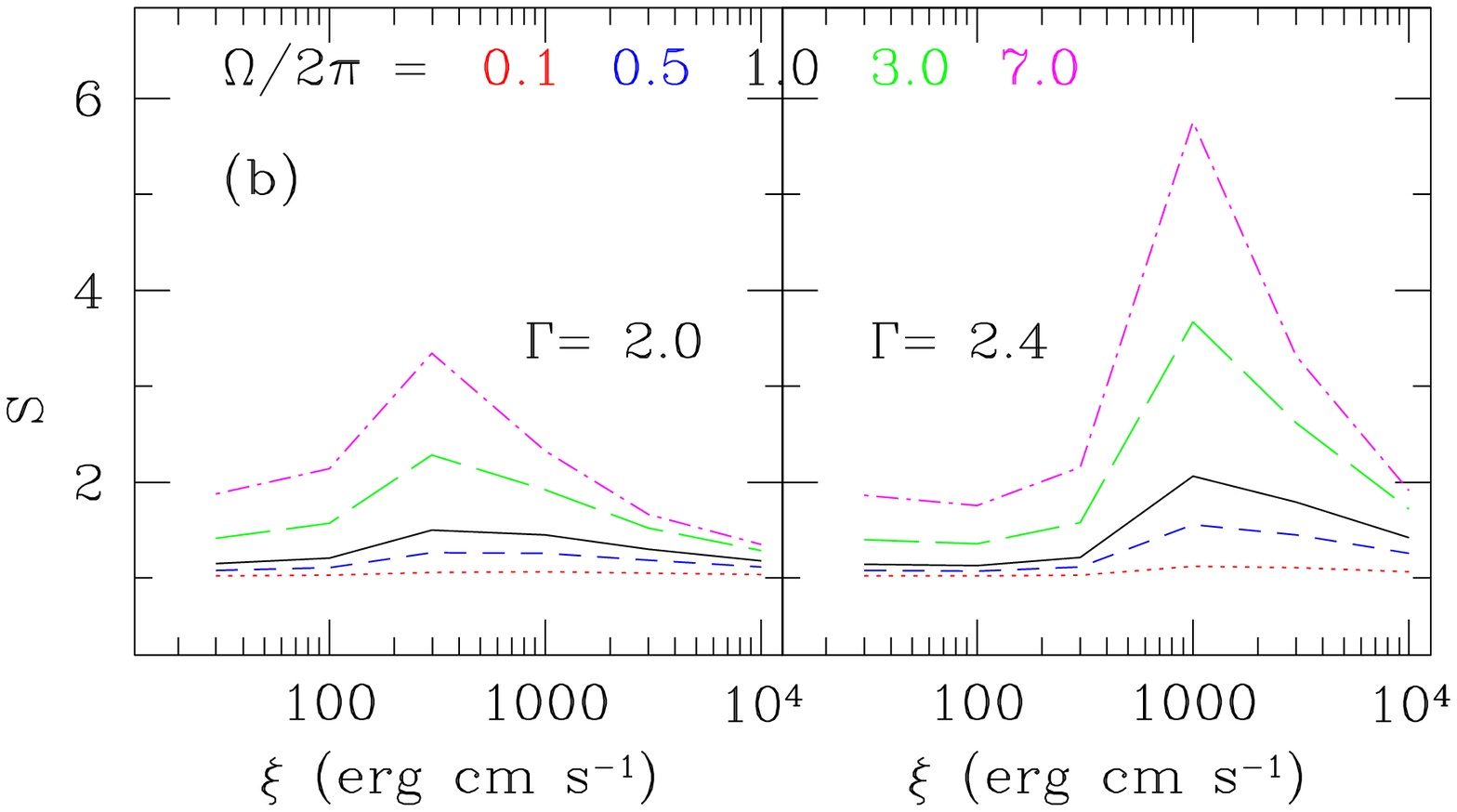}
\caption{Strength of the soft excess in a reflection model (with a
photon index of underlying continuum of $\Gamma = 2.0$ and $2.4$) as a
function of ionisation parameter and (a) iron abundance for
$\Omega/2\pi = 1$, (b) amplitude of reflection for Solar
abundance. The curves correspond to Fe abundance (or $\Omega/2\pi$) of
0.1 (dotted), 0.5 (short-dashed), 1 (solid), 3 (long-dashed), and 7
(dot-dashed).}  
\label{fig:sereflion}
\end{figure*}

The structure of the paper is as follows. In Sec.~2 we try to
distinguish between the reflection and absorption 
models for the soft excess. We calculate the strength of the soft
excess in both models, showing that the largest observed soft excesses
can be produced in the absorption model by simply changing the column
of material in the line of sight.  In Sec.~3 we explicitly
compare reflection and absorption models by fitting data from
PG~1211+143 and 1H~0707-495, representative of the largest soft
excesses seen in Quasars and NLS1's, respectively. Both models give
comparably good fits to the data in terms of $\chi^2$ criteria, as
does the more phenomenological model of a separate soft
component. Statistics cannot discriminate between them, and depending
on $\chi^2$ alone may be somewhat misleading as there are systematic
uncertainties in the models (such as range of ionisation states
present) which probably dominate the residuals.  

How then can we distinguish between them? The {\it XMM--Newton} data
is fairly high signal--to--noise, moderate spectral resolution and
fairly broad bandpass (0.3--10~keV), yet neither energy nor
variability spectra can clearly discriminate between these
models. In Sec.~4 we suggest that simultaneous data at higher
energies may be able to break these spectral model degeneracies. We
also review the physical plausibility of the models, which leads us to
favour the smeared absorption as the origin of the soft excess.

\section{Comparison of reflection and absorption models for the soft
  excess}

We use the results of GD04 to determine the range of soft excess
strengths which are required by the data. They defined $R_x$ as the
ratio of unabsorbed 0.3--2~keV fluxes in the soft excess and power law
continuum components, respectively. However, the data need to be
fitted in order to use such parametrisation.  An alternative measure of
the size of soft X-ray excess, which does not require spectral
fitting, is to assume that the spectrum is a power law with single 
photon index $\Gamma$ but with a step in normalisation at 0.7~keV,
i.e. $f(E)=AE^{-\Gamma+1}$ for $E\ge0.7$~keV, and $f(E)=S AE^{-\Gamma+1}$
for $E<0.7$~keV. The ratio $S$ of the spectrum at soft energies to
that expected from an extrapolation of the higher energy power law is
often plotted in the literature (e.g. Porquet et al. 2004), and
relates rather simply to the soft excess parameterisation of GD04 by
$R_x=a(\Gamma)(S-1)+1$ where $a(\Gamma) =
(0.7^{2-\Gamma}-0.3^{2-\Gamma})/(2^{2-\Gamma}-0.3^{2-\Gamma})=
0.3$--0.6 for $\Gamma=1.5$--2.5. Thus with this parametrisation we can
use the objects with ratio plots in the literature as well as data
from the sample of GD04. These range from 2--3 for the majority of the
PG Quasars (GD04; Porquet et al. 2004) to $\ga$10 for some
NLS1's (Boller et al. 2002, 2003; Reeves et al. 2004).

In reflection models, the size of the soft X-ray excess will be set by
the solid angle subtended by the reflecting material, and its
ionisation state. We use ionized reflection spectra of Ross \& Fabian (2005) 
publicly available as {\sc atable} models in {\sc xspec}. We fix the
the inclination at 30$^\circ$, and smear the
model by relativistic effects so that the line features do not
dominate (we fix the reflector emissivity at $3$, and its inner and
outer radius at 1.24$R_g$, i.e. extreme Kerr, and 400$R_ g$, respectively).

Figure~\ref{fig:sereflion}a shows how the spectral ratio $S$ (evaluated
at 0.5~keV) depends on the ionisation parameter of the reflector for
the maximum 'normal' reflection amount of $\Omega/2\pi = 1$. Firstly,
it is apparent that partial ionisation is required in order to have
the soft excess.  The strong jump in opacity due to the transition
from the Oxygen K and iron L shell is only present when these species
are present, which requires ($2.5\le \log \xi \le 3.5$).  Secondly,
the strength of this soft excess is rather limited.  Allowing for
super Solar iron abundance does not change this conclusion (see
Fig.~\ref{fig:sereflion}a) as it is set by the solid angle of the
reflector. Even in the limit of perfectly ionized reflection below
0.7~keV and no reflection at 3~keV, this sets a maximum $S\sim
1+\Omega/2\pi<2$ for 'normal' reflected emission with $\Omega/2\pi
<1$. 

The majority of soft excesses seen from the PG Quasars do indeed have
$S\la$~2--3 (GD04; Porquet et al. 2004),
but the largest soft excesses seen require that the spectrum is
dominated by the reflected emission rather than the intrinsic
continuum (Fig.~\ref{fig:sereflion}b).  This could be done in a
geometry in which the disc fragments, or from lightbending distorting
the illumination pattern of an initially isotropic source
(Fabian et. al. 2002, 2004; Miniutti \& Fabian 2004). However, if the
lightbending interpretation is correct, this predicts a correlation
between the solid angle (reflection dominance of the spectrum) and the
amount of smearing as stronger lightbending also focuses the
illumination more strongly onto the inner disc. Such a correlation is
not seen in the data (Crummy et al. 2006).

\begin{figure}
\centering
\includegraphics[width=5.3cm,bb=18 144 320 444,clip]{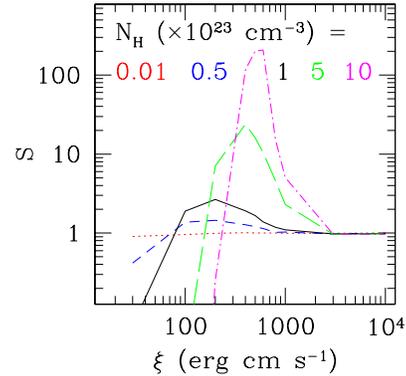}
\caption{Strength of the soft excess in an absorption model as a function of ionisation
parameter for column densities N$_H$ ($\times
10^{23}$ cm$^{-3}$) of 0.01 (dotted), 0.5 (short-dashed), 1 (solid),
5 (long-dashed), and 10 (dot-dashed).
The sharp absorption features were smeared with $\sigma=0.2$. In
this model the strength of the soft excess does not depend on
the photon index of the continuum.} 
\label{fig:sesmfix}
\end{figure}

Absorption models also require a range of ionisation parameter
as again they are dependent on the strong opacity jump produced by
Oxygen K/iron L transitions.  To quantify the soft excess strength in
the smeared absorption model we assume that the absorber has a gaussian
velocity distribution with $\sigma=0.2c$ modifying an intrinsic power
law continuum with photon index of $\Gamma = 2$. The soft excess is
then defined as the ratio of the absorption model to extrapolation of
a power law fit to the 3--8~keV spectrum at 0.5~keV, and its strength
as a function of ionisation parameter is shown in
Fig.~\ref{fig:sesmfix}. The soft excess disappears for $\log \xi \la  
2$ as the neutral material absorbs all the soft X-ray emission, and
for $\log \xi \ga 4$ where the column is completely ionized and
transparent at all energies. However, the soft excess at $\log \xi
\sim 3$ is now strong enough to match the largest seen. (Note the
increase in y--axis range of this figure compared to the reflection
plots, and the log scale.) Changing the column changes the size of
soft excess, as shown by Fig.~\ref{fig:sesmfix}.


Both the reflection and absorption models require a 'fine-tuned'
ionisation parameter, such that Oxygen is partially ionized. This may
be explained in the absorption models by the ionisation instability,
which results from X-ray illumination of material in some sort of
pressure balance (Chevallier et al. 2006). However, there is no
obvious way to do this in reflection models. The ionisation
instability in a disc {\em reduces} the extent of the partially
ionized zone (Nayakshin, Kazanas \& Kallman 1999), yet to produce a soft excess
from reflection requires that the partially ionized zone has to
dominate over the entire photoshere of the disc.  It is very difficult
for a hydrostatic disc to produce the soft excess (Done \& Nayakshin,
2006, in preparation). Even without the response of the disc to the
irradiating flux, X-ray illumination of a standard
Shakura-Sunyaev disc gives an ionisation
parameter
\begin{eqnarray*}
\xi (r) & = & L_X/n(r) (H^2+R^2) \\
        & = & 2 \times 10^{11}\, f\,  \dot{m}^3\, \alpha\, r^{-1.5}\, (h^{2}+r^2)^{-1}\, c_3^{-1}\,\, {\rm erg\, cm\, s^{-1}},
\end{eqnarray*}
where disc density
$n(r)=2.4{\times}10^7\alpha^{-1}\dot{m}^{-2}m_9^{-1}r^{1.5}c_3$~cm$^{-3}$ is
taken from Laor \& Netzer (1989), $\alpha$ is the viscosity parameter,
$\dot{m} = L/L_{\rm Edd}$, $L_{\rm Edd} =
1.3{\times}10^{47} m_{\,9}$ ergs s$^{-1}$, $m_{\,9} = M/(10^9\, {\rm
M_{\odot}})$, $L_X = f L$, so that $f$ is the fraction of
the luminosity which goes into hard X-rays, $c_3$ is a relativistic
correction, and all the distances are expressed in
terms of gravitational radius, $R_g = 1.5{\times}10^{14}m_{\,9}$,
i.e. $R = r R_g$, $H = h R_g$. A similar relation is
given by Ross \& Fabian (1993). This can give the required value of
$\xi \sim 10^{2-3} {\rm erg\, cm\, s^{-1}}$, but is very dependent on
the parameters, changing particularly rapidly with $\dot{m}$. 
An alternative
prescription for the viscosity, where the heating goes as the
geometric mean of the gas and radiation pressure (which may be more
physically realistic: Merloni 2003), results in a denser
disc and hence a lower ionisation parameter of
\begin{eqnarray*}
\xi (r) = ~~~~~~~~~~~~~~~~~~~~~~~~~~~~~~~~~~~~~~~~~~~~~~~~~~~~~~~~~~~~~~~~~~~~~~& &\\
8\times 10^6 f\, \dot{m}^{2.11}\, m_{\,9}^{-0.12}\, \alpha^{0.88}\,
(h^{2}+r^2)^{-1}\, r^{-0.33}\, c_7^{-1}\,\,{\rm erg\, cm\, s^{-1}},& &
\end{eqnarray*}
where $c_7$ is a relativistic correction. This is less strongly
dependent on parameters, but both require some method of 
'fine-tuning' the source height relative to $\dot{m}$ in order to
produce the observed narrow range in ionisation parameter.


\begin{figure*}
\centering
\includegraphics[width=88mm]{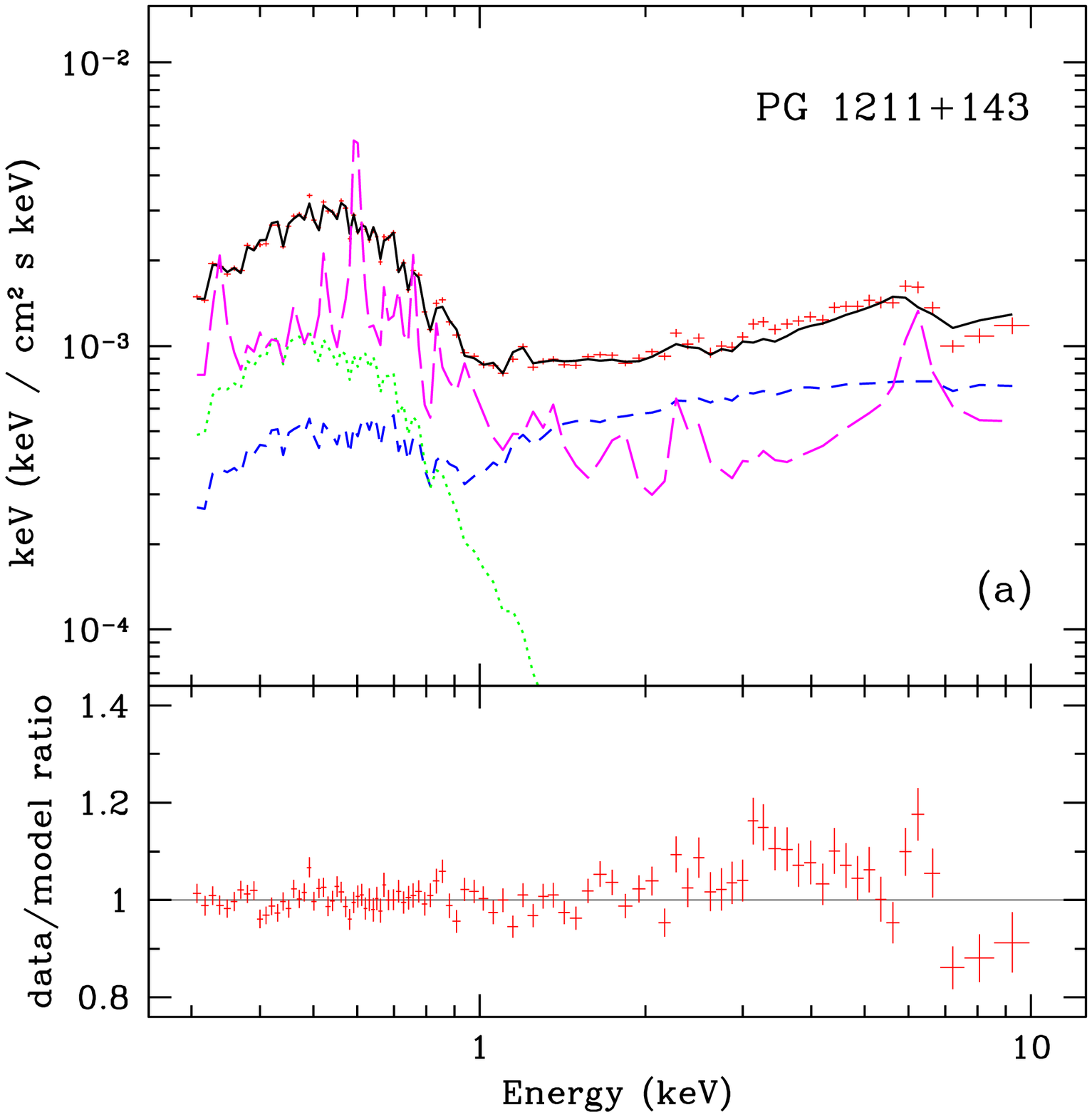}
\includegraphics[width=88mm]{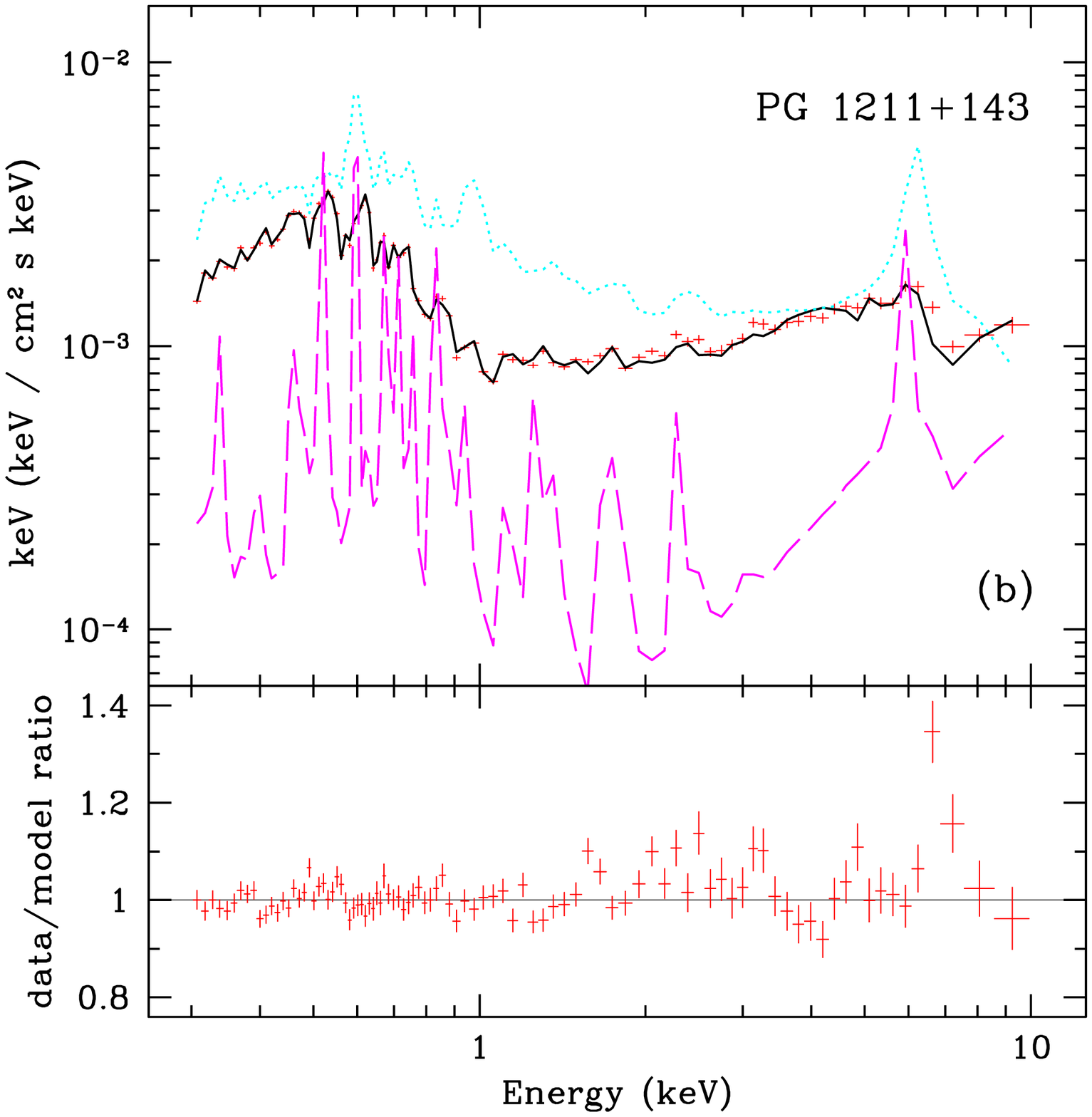}\\
\includegraphics[width=88mm]{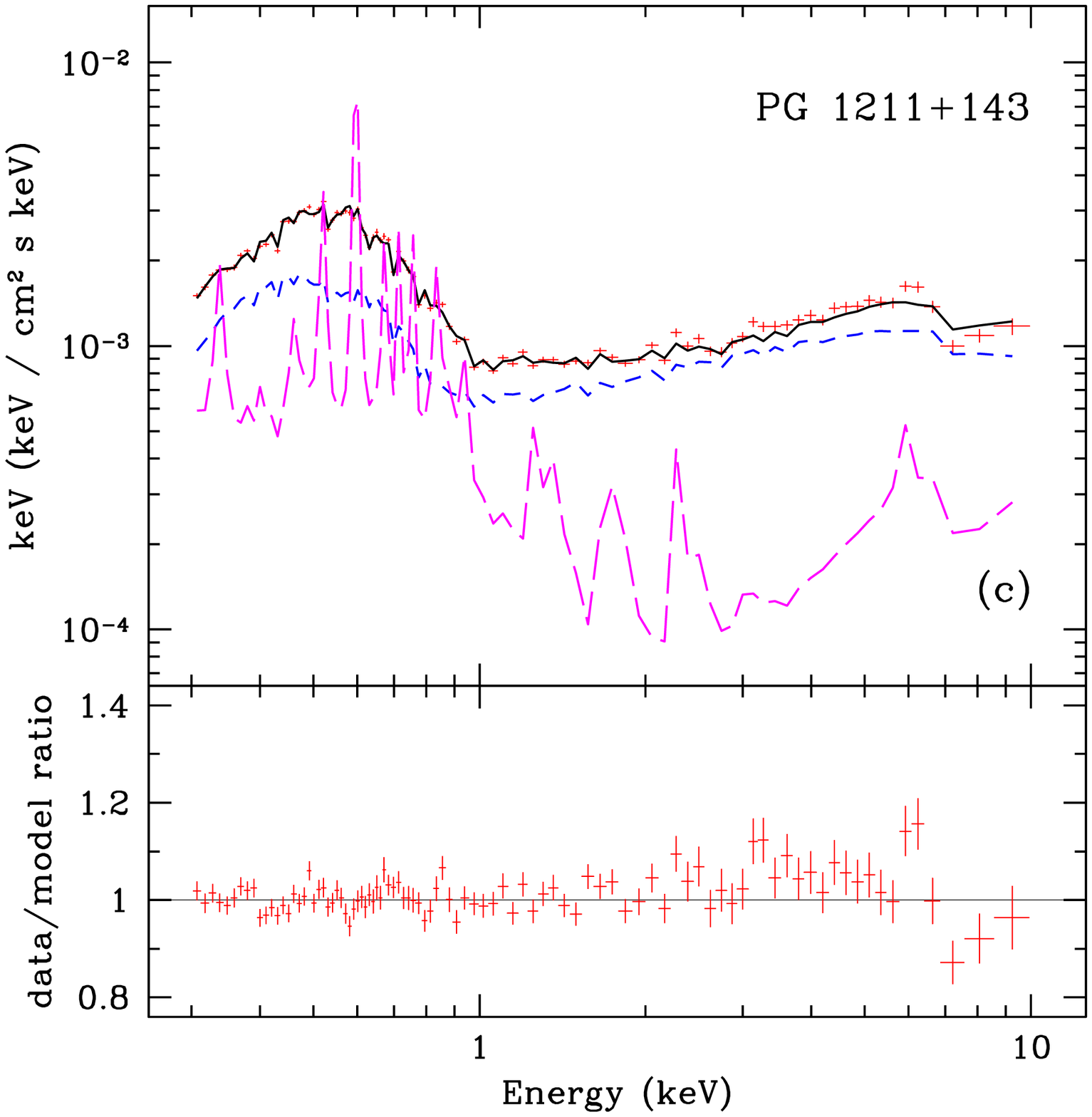}
\includegraphics[width=88mm]{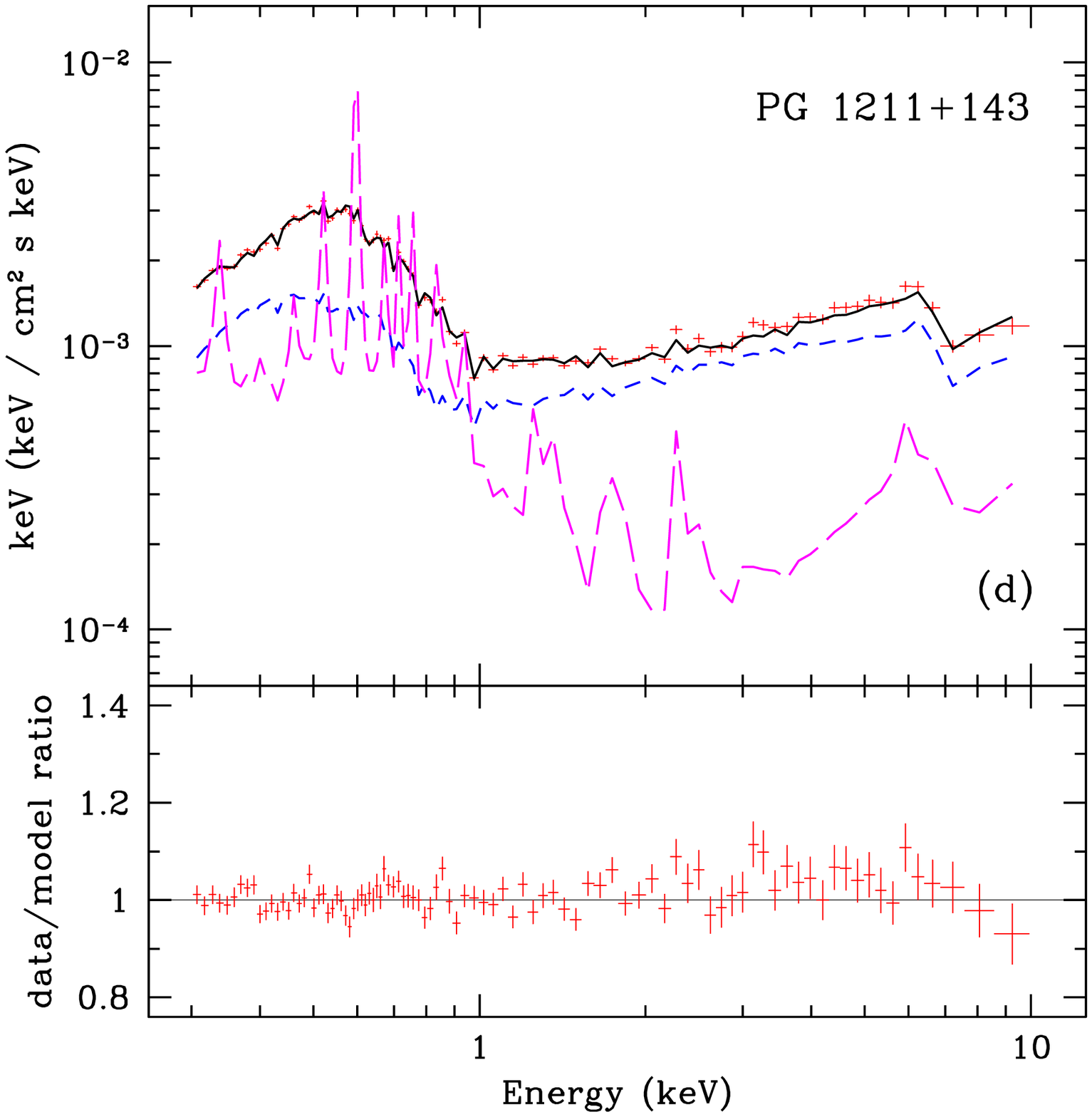}
\caption{Modelling the soft excess in PG~1211+143. Solid curve: the best fit
spectra. Model~1 (a) -- a low temperature Comptonization (green/dotted), power
law (blue/short dashed) and ionized reflection (magenta/long
dashed). Model~2 (b) -- two reflectors (magenta/dotted and cyan/long dashed) with
different column densities and ionization states, located at different
radii. Model~3 (c) -- a power law (blue/short dashed) subject to
relativistically smeared absorption and reflection (magenta/long
dashed). Model~4 (d) -- same as Model~3 but with the intrinsic power
law modified by the addition of a line with P Cygni profile modelling
the iron feature.  All four models are additionally affected by
galactic absorption, cold absorption at the redshift of the sources,
and one or two warm absorbers (see text).}
\label{fig:pg}
\end{figure*}

\begin{figure*}
\centering
\includegraphics[width=88mm]{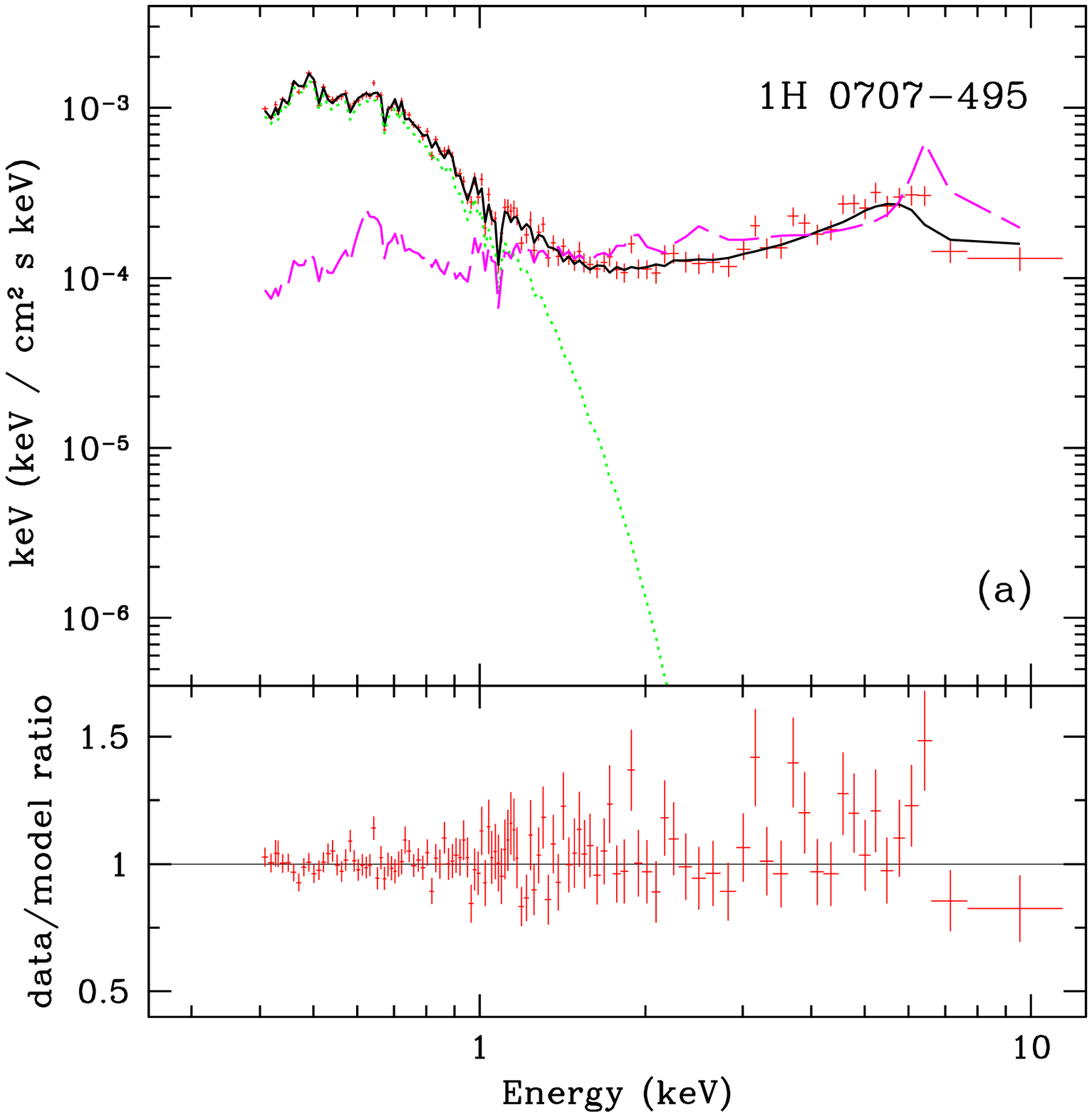}
\includegraphics[width=88mm]{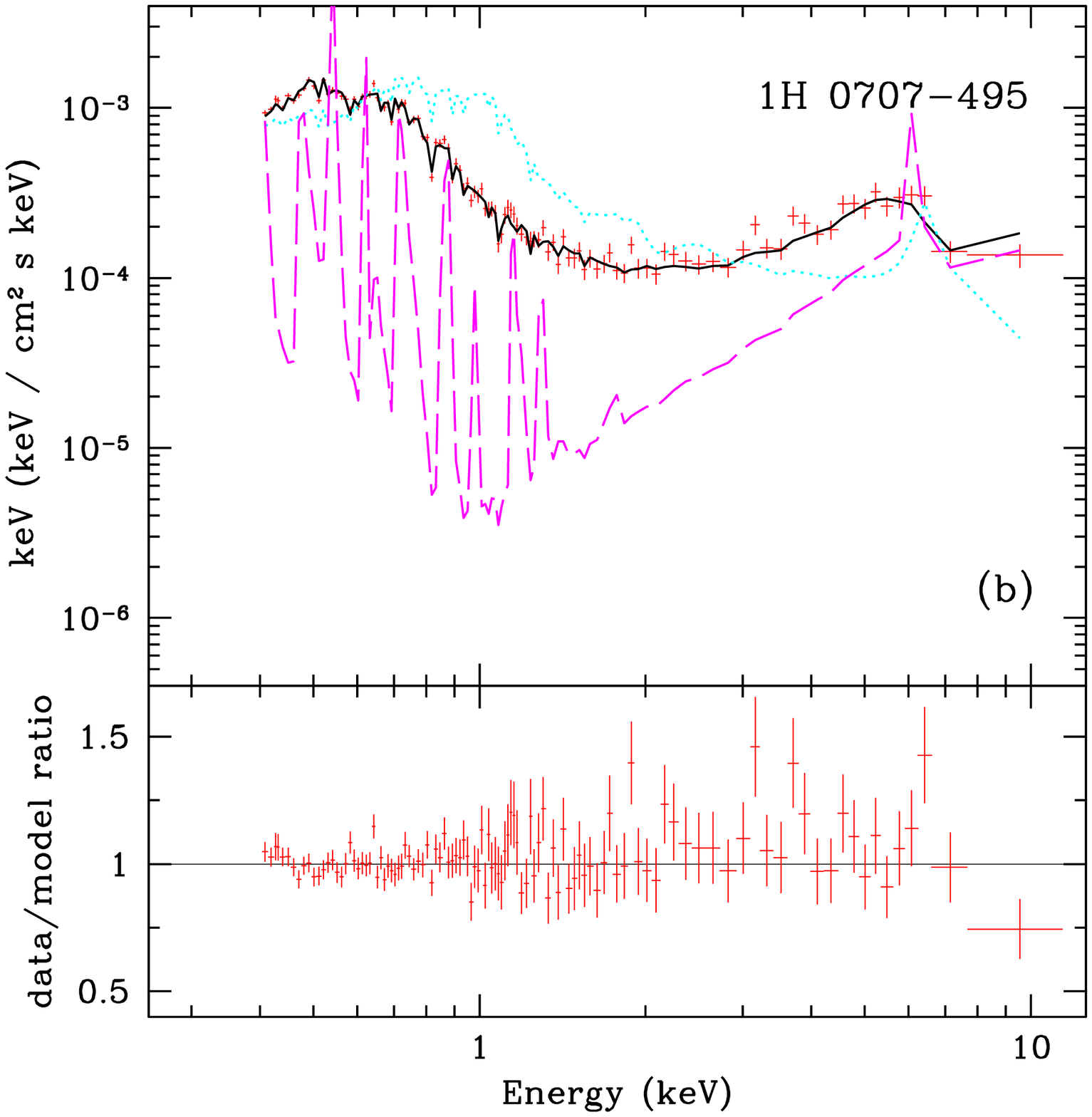}\\
\includegraphics[width=88mm]{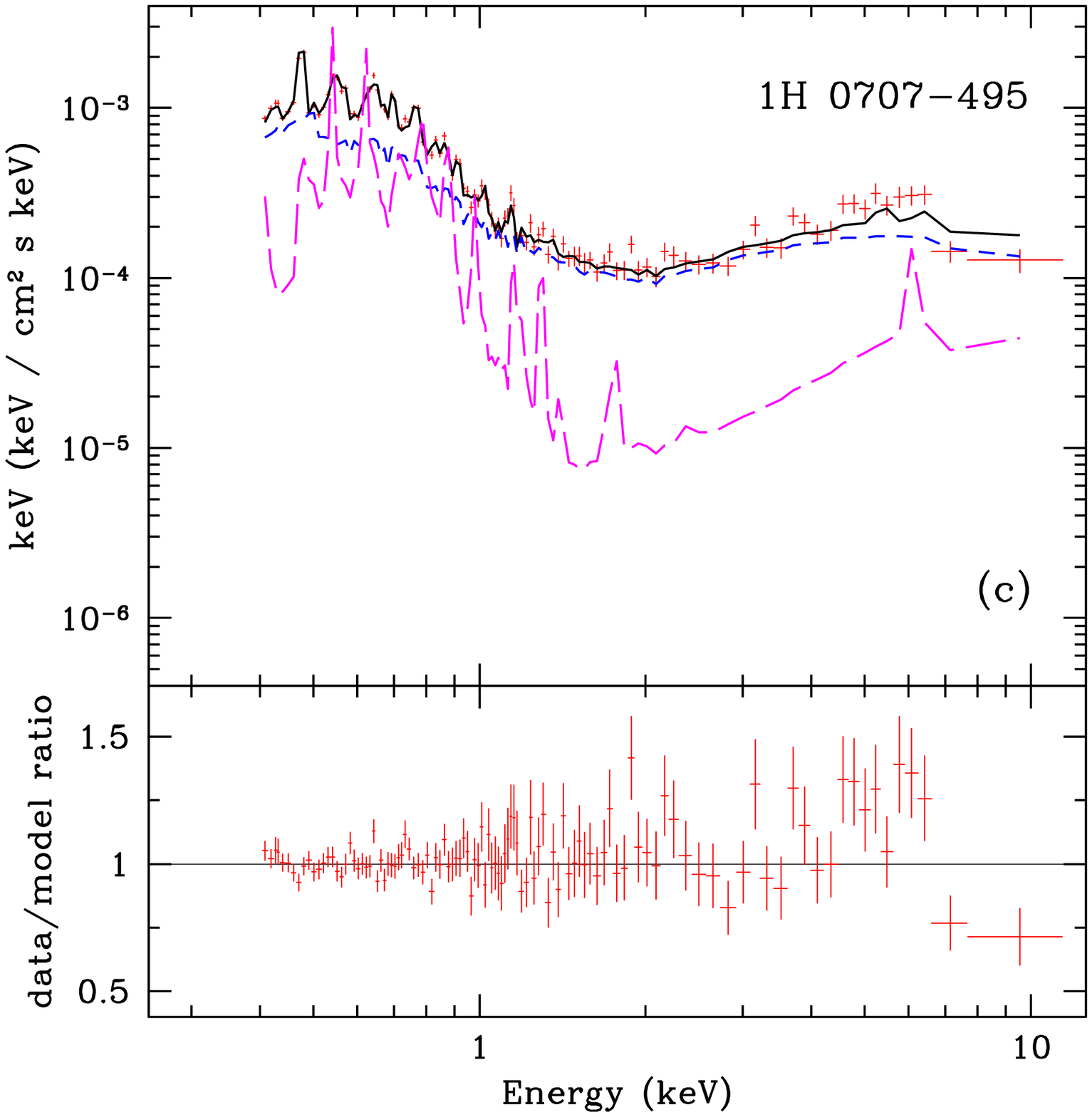}
\includegraphics[width=88mm]{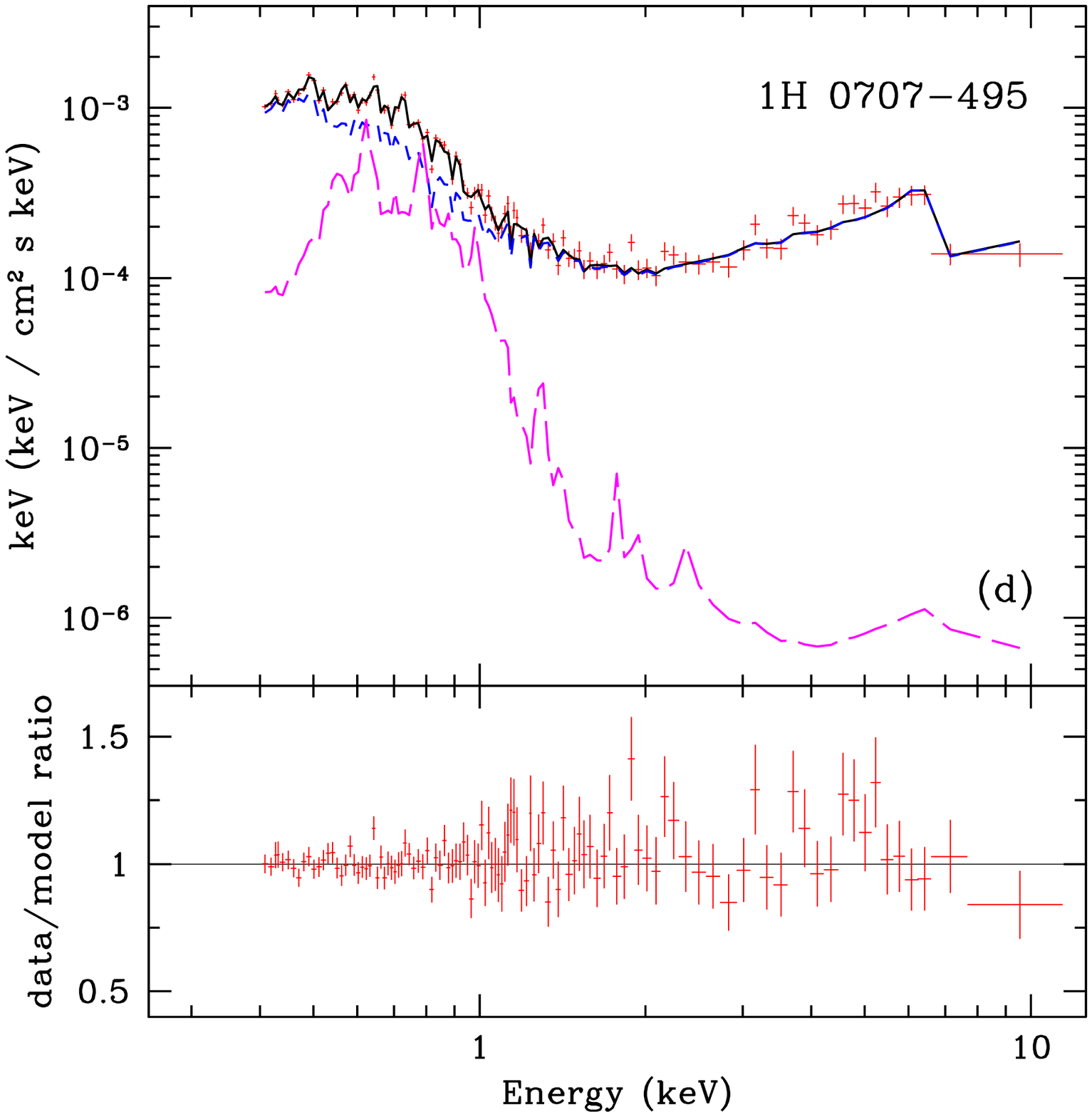}
\caption{Modelling the soft excess in 1H~0707-495. (a), (b), (c), (d)
as in Fig.~\ref{fig:pg}. Model~1 (a) in 1H~0707-495 does not require
intrinsic power law.}
\label{fig:1h}
\end{figure*}


\begin{table*}
\caption{Best fit parameters of Model 1 to PG~1211+143 and 1H~0707-495
data. Plasma optical depth was fixed at $\tau=50$, emissivity at 3 and
inclination at 30$^{\circ}$. No intrinsic power law emission is seen
in the model for 1H~0707-495.
} 
\vspace{1.5mm}
\centering
\begin{tabular}{l c c c c c c}
\hline
~ & $kT_e$ & $\Gamma$ & $\log \xi$ & $R_i$ & $R_f$ & $\Omega/2\pi$ \\  
~ & keV & & & $R_g$ &$R_g$ & \\
\hline
PG & 0.136$\pm 0.003$ & 2.02$^{+0.03}_{-0.02}$
   & 2.58$^{+0.04}_{-0.01}$ & 
4.5$^{+0.7}_{-0.5}$ & 8.7$^{+1.8}_{-2.2}$ & 4.12$\pm 0.49$ \\  
1H & 0.122$ \pm 0.002$ & 1.81$^{+0.07}_{-0.09}$
   & 3.00$^{+0.02}_{-0.15}$ & 
3.0$^{+1.0}_{-0.5}$  & 7.4$^{+4.5}_{-1.8}$ & $\infty$ \\ 
\hline
\end{tabular}
\label{tab:m1}
\end{table*}

\begin{table*}
\caption{Best fit parameters of Model 2 to PG~1211+143 and 1H~0707-495
data. Missing upper (lower) limit indicates that the parameter
reached its upper (lower) boundary allowed in the fit. Emissivity was
fixed at 3 and inclination at 30$^{\circ}$ and the iron abundance was
twice the Solar value. The fits are reflection dominated, so
$\Omega/2\pi = \infty$.
}
\vspace{1.5mm}
\centering
\begin{tabular}{l c ccc ccc}
\hline
~ & $\Gamma$ & $\log \xi_1$ & $R_{i_1}$ &$R_{f_1}$ & 
$\log \xi_2$ & $R_{i_2}$ & $R_{f_2}$ \\
~ & & & $R_g$ & $R_g$ & & $R_g$ & $R_g$ \\   
\hline
PG & 2.127$^{+0.006}_{-0.005}$ & 
3.13$\pm 0.01$ & 1.31$^{+0.21}_{-0.07*}$ & 3.20$^{+0.10}_{-0.06}$ & 
1.88$\pm 0.02$ & 22$^{+3}_{-8}$ & 33$^{+8}_{-4}$ \\
1H & 2.592$^{+0.009}_{-0.013}$ & 
4.00$_{-0.04}$ & 3.4$\pm 0.7$ & 4.4$^{+1.4}_{-1.5}$ & 
1.5$^{+0.2}$ & 1.89$^{+0.44}_{-0.65*}$  & 20$^{+16}_{-8}$ \\
\hline
\end{tabular}
\label{tab:m2}
\end{table*}

\begin{table*}
\caption{Best fit parameters of Model 3 to PG~1211+143 and 1H~0707-495
data. An asterisk indicates that the upper (lower) limit allowed was reached
with $\Delta \chi^2 < 2.71$. Emissivity was fixed at 3 and inclination at 30$^{\circ}$.} 
\vspace{1.5mm}
\centering
\begin{tabular}{l c ccc ccc c}
\hline
~ & $\Gamma$ & N$_H$ & $\log \xi_a$ & $\sigma$ & 
$\log \xi_r$ & $R_i$ & $R_f$ & $\Omega/2\pi$ \\
~ & & 
$\times 10^{22}{\rm cm}^{-2}$ & & & & $R_g$ & $R_g$ & \\   
\hline
PG & 2.34$^{+0.03}_{-0.05}$ & 
5.3$^{+0.8}_{-0.6}$ & 2.43$^{+0.04}_{-0.06}$ & 0.10$\pm 0.01$ & 
2.48$^{+0.04}_{-0.07}$ & 4.8$\pm 0.7$ & 12$^{+6}_{-3}$ &
1.67$\pm 0.47$\\
1H & 2.994$^{+0.006*}_{-0.034}$ & 
41$^{+9}_{-13}$ & 2.89$^{+0.06}_{-0.08}$ & 0.36$^{+0.09}_{-0.07}$ &
2.478$\pm 0.001$ & 32.4$^{+0.3}_{-2.0}$ & 33.8$\pm 0.5$ &
1.59$\pm 0.83$\\
\hline
\end{tabular}
\label{tab:m3}
\end{table*}

\begin{table*}
\caption{Best fit parameters of Model 4 to PG~1211+143 and 1H~0707-495
data. An asterisk indicates that the upper (lower) limit allowed was reached
with $\Delta \chi^2 < 2.71$.  Emissivity was fixed at 3 and inclination at 30$^{\circ}$.
} 
\vspace{1.5mm}
\centering
\begin{tabular}{l c ccc ccc ccc c}
\hline
~ & $\Gamma$ & 
$N_H$ & $\log \xi_a$ & $\sigma$ & 
$E$ & $v_{\infty}/c $ & $\tau$ &
$\log \xi_r$ & $R_i$ & $R_f$ & $\Omega/2\pi$ \\
~ & & 
$\times 10^{22}{\rm cm}^{-2}$ & & &
keV & & & $R_g$ & $R_g$ & \\   
\hline
PG & 2.29$^{+0.07}_{-0.05}$ & 
5.0$^{+1.1}_{-0.6}$ & 2.43$^{+0.06}_{-0.05}$ & 0.09$\pm 0.01$ &
6.58$^{+0.08}_{-0.10}$ & 0.5$_{-0.2}$ & 0.23$^{+0.19}_{-0.09}$ & 
2.49$^{+0.04}_{-0.13}$ & 4.4$^{+0.7}_{-0.6}$ & 12$^{+6}_{-2}$ &
1.95$\pm 0.65$\\
1H & 2.90$^{+0.07}_{-0.19}$ & 
48$^{+2*}_{-11}$ & 2.93$^{+0.05}_{-0.19}$ & 0.43$^{+0.07*}_{-0.06}$ &
6.58$\pm 0.17$ & 0.5$_{-0.1}$ & 2.0$^{+4.0}_{-1.1}$ &
3.12$^{+0.15}_{-0.08}$ & 60$^{+5}_{-11}$ & 66$^{+5}_{-2}$ &
0.011$\pm 0.002$\\
\hline
\end{tabular}
\label{tab:m4}
\end{table*}

\begin{table*}
\caption{Best fit parameters for warm absorbers (with $v_{\rm
    turb}~=~200$~km~s$^{-1}$) and cold absorption at the redshift of
    PG~1211+143 and 1H~0707-495. Missing upper (lower) limit indicates
    that the parameter reached its upper (lower) boundary allowed in
    the fit. An asterisk indicates that the upper (lower) limit allowed was reached
    with $\Delta \chi^2 < 2.71$.  Emissivity was fixed at 3 and inclination at 30$^{\circ}$.} 
\vspace{1.5mm}
\centering
\begin{tabular}{l c ccc ccc}
\hline
~ & $N_{H_1}$ (cm$^{-2}$)& $\log \xi_1$ & $z_1$ & 
$N_{H_2}$ (cm$^{-2}$)& $\log \xi_2$ & $z_2$ & 
$N_{H_z}$ (cm$^{-2}$)\\ 
\hline
~ & \multicolumn{7}{c}{PG} \\
\hline
Model 1 & (5.3$\pm 0.4$)$\times 10^{21}$ &
1.57$^{+0.03}_{-0.04}$ & -0.015$^{+0.001}_{-0.002}$ &
3.2$^{+1.5}_{-0.6} \times 10^{22}$ & 2.79$\pm 0.03$ &
-0.087$^{+0.004}_{-0.001}$ & 6.6$^{+3.7}_{-2.7} \times 10^{19}$ \\  
Model 2 & (4.6$\pm 0.3$)$\times 10^{21}$ & 
1.65$^{+0.03}_{-0.07}$ & 0.023$^{+0.003}_{-0.005}$  & 
(1.7$\pm 0.3$)$\times 10^{23}$  & 3.24$^{+0.09}_{-0.05}$ & 
-0.047$^{+0.003}_{-0.002}$ & $< 2.1 \times 10^{19}$  \\
Model 3 & 5.7$^{+3.0}_{-2.3} \times 10^{22}$ & 
2.90$\pm 0.04$ & -0.148$^{+0.003}_{-0.002*}$ & 
1.4$^{+0.5}_{-0.3} \times 10^{23}$ & 2.89$^{+0.07}_{-0.04}$ & 
-0.059$^{+0.003}_{-0.002}$ & 1.8$^{+0.6}_{-1.1} \times 10^{20}$ \\
Model 4 & 4.4$^{+4.9}_{-1.9} \times 10^{22}$ & 
2.82$\pm 0.04$ & -0.065$^{+0.007}_{-0.011}$ & 
5.9$^{+4.6}_{-2.0} \times 10^{22}$ & 2.84$^{+0.05}_{-0.04}$ & 
-0.150$^{+0.002}$ & 1.1$^{+0.8}_{-0.5} \times 10^{20}$ \\
\hline
~ & \multicolumn{7}{c}{1H}\\
\hline
Model 1 & 1.5$^{+2.4}_{-0.5*} \times 10^{21}$ &
 1.1$^{+0.4}_{-0.1*}$ & -0.02$\pm 0.04$ & 
9.4$^{+1.5}_{-1.9} \times 10^{21}$ & 1.73$^{+0.07}_{-0.05}$ & 
-0.148$^{+0.006}_{-0.002*}$ & (4.4$^{+1.4}_{-1.5}$)$ \times 10^{20}$ \\
Model 2 & 4.8$^{+1.5}_{-1.3} \times 10^{21}$ & 
1.5$\pm 0.1$ & -0.02$\pm 0.02$ & 
1.0$_{-0.7} \times 10^{24}$ & 3.42$^{+0.24}_{-0.05}$ & 
-0.06$^{+0.01}_{-0.02}$ & $< 1.6 \times 10^{20}$ \\
Model 3 & 2.0$^{+1.8}_{-1.0*} \times 10^{21}$ & 
1.5$^{+0.4}_{-0.5}$ & -0.02$^{+0.02}_{-0.03}$ & 
3.6$^{+6.4*}_{-2.7} \times 10^{23}$ & 3.31$^{+0.10}_{-0.06}$ & 
-0.04$^{+0.02}_{-0.01}$ &  6.3$^{+1.1}_{-0.7} \times 10^{20}$ \\
Model 4 & 4.0$^{+1.5}_{-2.4} \times 10^{21}$ & 
1.5$^{+0.1}_{-0.5*}$ & -0.02$^{+0.02}_{-0.03}$ & 
4.7$^{+1.1}_{-0.9} \times 10^{23}$ & 3.20$^{+0.08}_{-0.07}$ & 
-0.05$^{+0.02}_{-0.01}$ & $< 5.9 \times 10^{20}$ \\
\hline
\end{tabular}\\
\label{tab:abs}
\end{table*}

\section{Spectral fitting}

The theoretical considerations in the previous section favour an absorption 
origin for the soft excess as it can reproduce the observed range of
soft excess 
strengths without extreme geometries, and the required
'fine-tuning' of the ionisation parameter may have an explanation in 
terms of ionisation instability. However, given the model 
uncertainties, observations may be a better way to distinguish
between reflection and absorption as the origin of the soft excess. 

We use {\it XMM--Newton} data from two representative objects with large
soft excesses, the low-redshift ($z = 0.085$) quasar, PG~1211+143, and
the Narrow Line Seyfert 1 galaxy 1H~0707-495 ($z=0.0411$). PG~1211+143
is one of the extensively studied objects from Palomar-Green Bright
Quasars Survey and was selected based on its optical brightness
suggesting super-Eddington accretion rate (GD04; Boroson
2002). 1H~0707-495 is less luminous  
than PG~1211-143, but it shows one of the strongest soft excesses
seen. Both objects show a drop of the flux at 7 keV, which is
especially dramatic in the case of 1H~0707-495
(Boller et al. 2002). Their spectra are accretion disc dominated which
suggests that they are counterparts of X-ray binaries in the high/soft
state (Janiuk, Czerny \& Madejski 2001).

We use the Epic PN data, extracted from the database using standard 
techniques (see GD04). We fit these data in the 0.3--10 keV
energy band using {\sc xspec} to explicitly compare the 
models for the soft excess. 

Firstly we set up a baseline model where the soft excess is a separate
component. We use the {\sc comptt} code of Titarchuk (1994) to
describe the soft excess as Comptonization of disc photons (assumed to be at
fixed $T_{\rm bb}=10$~eV) on electrons of temperature $kT_e$,
and the optical depth, $\tau$ (Model~1). The soft excess shape
generally cannot easily constrain these two parameters separately, so
we fixed $\tau = 50$. The hard X-ray tail was modelled by a power law
(with photon index $\Gamma$ and normalisation) together with
its reflection from the accretion disc.  We use the publicly
available models of Ballantyne, Iwasawa \& Fabian (2001), which were then
relativistically smeared using the Laor (1991) kernel. This
reflection is characterized by the ionisation parameter of the
reflecting matter, $\xi$, and its normalisation, while the smearing
parameters are the inner and outer radii of the reflecting disc,
together with its inclination (fixed at 30 degrees). We fix the emissivity
at 3, which results in rather small value of the outer radius of the
reflector. Conversely, fixing the outer radius at (say) 400$R_g$ would
result in an emissivity greater than 3.

This baseline model is then modified to describe different potential
origins for the soft excess. In a reflection origin (Model~2) we
replace the {\sc comptt} component (which had 2 free parameters,
namely electron temperature and normalisation) with another
relativistically smeared reflector (4 additional parameters:
ionisation, normalisation, inner and outer radii), while in Model~3 we
replace it with the relativistically smeared absorption (the {\sc
  swind} model of Gierli\'nski \& Done 2006 which has 3 additional
parameters: ionisation, column and Gaussian velocity smearing,
$\sigma$).

All models were additionally modified by the cold Galactic absorption
(the {\sc wabs} model with column densities fixed at 2.7$\times 10^{20}$
cm$^{-2}$ and 5.8$\times 10^{20}$ cm$^{-2}$ for PG~1211+143 and
1H~0707-459, respectively), cold absorption at the redshift of the
source (the {\sc zwabs} model with $z=0.085$ and $z=0.0411$ for
PG~1211+143 and 1H~0707-459, respectively) and warm absorption
accounting for the narrow features in the spectra, modelled with the {\sc
xstar} code (Bautista \& Kallman 2001). We use the most recent {\sc
xstar} table models which are publicly available (grid 25) computed
for turbulent velocity of 200 km s$^{-1}$ and valid for wide range of
ionisation parameter and column density ($1 < \log \xi < 5$, $21 <
\log N_H < 24$). Hence, the only thing that differentiates the
models is the description of the origin of the soft excess.

The spectral decomposition implied by each model is shown in Figs.~\ref{fig:pg}
and \ref{fig:1h} for PG~1211+143 and 1H~0707-459, respectively. The high
resolution data have been rebinned for clarity of the plots. The
detailed model parameters are given in Tables
\ref{tab:m1}--\ref{tab:m4}. These show that the soft excess, if
described by Model~1, originates in the Comptonization of disc photons
on electron plasma with temperature of $kT_e \simeq 0.12$--0.14
keV. The power law modelling the high energy part of the spectrum is
relatively hard, $\Gamma \simeq 1.8$--2, and it is 
reflected by an ionized material with $\log \xi \simeq 2.6$--3.0, in the
innermost parts of the accretion flow (at 5--9$R_g$ and 3--7$R_g$ for
PG~1211+143 and 1H~0707-495, respectively). The last 
stable circular orbit in the Schwarzschild metric around a
non-rotating black hole is located at 6$R_g$, so neither of
these significantly requires a rotating Kerr black hole, though the
correspondingly small outer radii indicate that the emissivity is
highly centrally concentrated in both AGN. The amount of reflection as
compared to the intrinsic power law emission implies a solid angle
of $\Omega/2\pi \sim 4$ for PG~1211+143. For 1H~0707-459
the best fit has no intrinsic power law emission, so that the high
energy spectrum is formed solely by reflection. In both AGN the reflection
is partially ionized, so that around half of the soft excess is made
by the reflected component. However, this single reflector alone
cannot describe the shape of the spectrum; $\chi^2$ increases by 56
(940 d.o.f. in the model with {\sc comptt}) and 175 (286 d.o.f in the
model with {\sc comptt}) in PG~1211+143 and 1H~0707-495, respectively,
if the Comptonization component is removed.

However, if instead of the {\sc comptt} component, another smeared
reflector is introduced (Model~2) then the soft excess can be well fit
by the combination of the two reflectors with different ionisation
states, one mostly neutral (with $\log \xi < 2$) and the other highly
ionized (with $\log \xi \simeq 3$--4). However, this leaves systematic
residuals around ${\sim}7$~keV. A better fit to 1H~0707-495 was
obtained by assuming that the iron abundance in both reflectors  was
twice the Solar value ($\Delta \chi^2 = 18$ for 284 d.o.f.). The
improvement was mainly due to the reduction in residua around the iron
feature. In PG~1211+143 the fits with the Solar and twice the Solar
iron abundances are of the same quality. Substantial relativistic
smearing (in the case of PG~1211+143 requiring an extreme Kerr black
hole) is required to model the data. The incident X-ray radiation has
a moderate photon index in PG~1211+143 ($\Gamma \simeq 2.1$) and is rather
soft in 1H~0707-495 ($\Gamma \simeq 2.6$) but is not seen in either
AGN. The best fit spectra are dominated by the two reflectors only,
implying a geometry in which the intrinsic power law emission is
strongly suppressed, so it makes a negligible contribution to the
observed radiation.   
 
Replacing the second reflector with smeared absorption gives rather
different results (Model~3). The remaining reflector subtends a solid
angle $\Omega/2\pi < 2$, much smaller than in the case of reflection
dominated model. It is partially ionized with $\log \xi \simeq 2.5$,
and fairly close to the black hole (5--12$R_g$ and $\sim$30$R_g$ for
PG~1211+143 and 1H~0707-495, respectively) but is not 
so strongly smeared as to give constraints on spin. The partially
ionized reflection does contribute to the soft excess (see also
Chevallier et al. 2006), though this component could also be
produced by emission from the smeared absorbing material
(Schurch \& Done 2006). This latter 
has ionisation parameter $\log \xi = 2.4$--2.9, high column
density, $N_H \simeq (5$--40)$\times10^{22}$ cm$^{-2}$ and large
velocity shear ($\sigma = v/c \simeq 0.1$ and 0.4 for PG~1211+143 and
1H~0707-495, respectively).  Such matter may be physically pictured as
some sort of accretion disc wind, which may be differentially rotating
as well as accelerating (and/or decelerating) radially. The assumed
Gaussian velocity dispersion is only a zeroth order approximation to
this complexity.

While the reflection model leaves some systematic residuals around the
iron line, these are much more prominent in the absorption
model. Especially for 1H~0707-495, the absorption model does not describe
the sharp drop around 7~keV in the spectrum. This is partially due to
the assumption of Gaussian velocity smearing in the 
absorption model. By definition this cannot produce sharp features.
Nonetheless, sharp features can be produced by a wind with strong
velocity shear. P Cygni profiles are an obvious example of this.
Done et al. (2006) show that the 7~keV feature in these data from
1H~0707-495 can be fit using a P Cygni profile from
emission/absorption/scattering of the He-- or 
H--like resonance iron K$\alpha$ line at 6.7 or 7.0~keV,
respectively.  We extend the smeared absorption model by including a P Cygni
profile to model the ${\sim}7$~keV feature (Model~4: two additional
free parameters being the absorption optical depth of the line and its
rest energy).  Thus, the incident power law is not only affected by
the smeared absorption but also has a P Cygni line as in
Done et al. (2006). 

In Model~4 applying the P Cygni profile to the ${\sim}7$ keV feature
results in a much improved fit around the iron feature. The inferred line 
rest frame line energy of 6.85--7.14~keV, which is
consistent with either He-- or H--like iron. The fits give $v_{\infty}/c
= 0.5_{-0.2}$ and $v_{\infty}/c = 0.5_{-0.1}$ for PG~1211+143 and
1H~0707-495, correspondingly. The resonance absorption line optical
depth in PG~1211+143 is relatively low, $\tau = 0.23^{+0.19}_{-0.09}$,
and in 1H~0707-495 it is higher and yields $\tau = 2.0^{+4.0}_{-1.1}$. 
 As in Model~3, the parameters of the remaining reflector are much
less extreme than for the reflection model of the soft excess
(Model~2). In fact, the deconvolved spectra in Figs.~\ref{fig:pg}d and 
\ref{fig:1h}d show that this reflector contributes mainly to producing 
soft X-ray lines rather than the iron feature, indicating that the
reflection parameters may be further distorted if there is also emission from
the wind (see e.g. Schurch \& Done 2006).

Table~\ref{tab:abs} contains parameters of the warm absorbers
modifying the continua such as column density, logarithm of
the ionisation parameter and redshift of the warm absorbers. We find
that including the second warm absorber significantly improves the fits in
 all cases. The last column in Table~\ref{tab:abs} contains the column
density of the cold absorption at the redshifts of the source. It is
comparable with the Galactic cold absorption for 
all models except Model~2 and Model~4 (the 1H~0707-495 case) for which
only the upper limit can be given. 

The best fit values of $\chi^2$ and number of degrees of freedom are
given in Table~\ref{tab:chi}. In PG~1211+143 the worst fit is obtained
with the reflection dominated model. The smeared absorption model is
statistically significantly better than the model with low temperature
Comptonization (F-test probability of chance improvement 5$\times
10^{-8}$). Modelling the iron feature with a P Cygni profile further
improves the fit (F-test probability $10^{-5}$). In
1H~0707-495 the smeared absorption model fits worse than either
reflection dominated model or the model with additional
Comptonization. However, including the P Cygni line in a fit improves
it dramatically, giving the best overall fit.  The F-test probability
of chance improvement in Model~4 as compared to Models~1 and 2 is
of the order of $10^{-4}$.

\begin{table}
\caption{The $\chi^2$ values and degrees of freedom in spectral fits to the
PG 1211+143 and 1H 0707-495 data.} 
\vspace{1.5mm}
\centering
\begin{tabular}{l c c c c}
\hline
~ &  Model 1 & Model 2 & Model 3 & Model 4\\
~ & ~ & $\chi^2$ (d.o.f.) & ~ \\
\hline
PG & 994 (940) & 1012 (939) & 963 (939) & 937 (936)\\
1H & 296 (286) & 291 (284)  & 314 (284) & 271 (281)\\
\hline
\end{tabular}
\label{tab:chi}
\end{table}


\begin{figure}
\centering
\includegraphics[width=88mm]{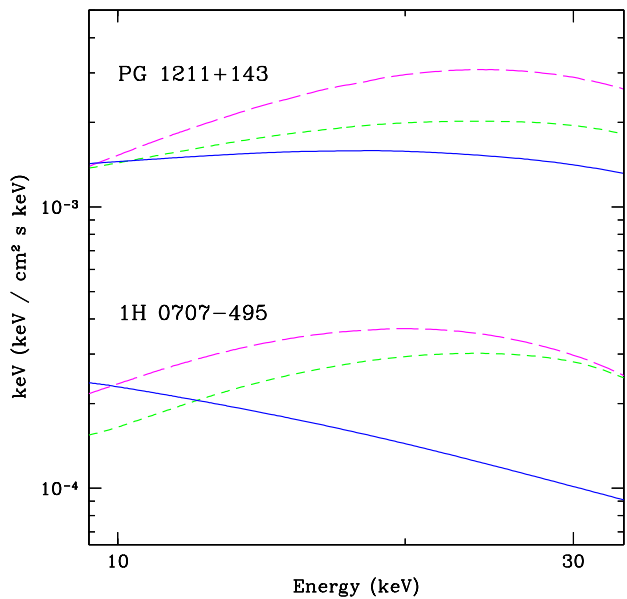}
\caption{Extrapolation of best fit model continua to the 10--30 keV
  range. The models gives different predictions on the 10--30 keV
  flux, with the reflection dominated model (magenta/long-dashed)
  producing the highest flux and the smeared absorption model
  (blue/solid) producing the lowest flux. The model with low
  temperature Comptonization (green/short-dashed) predicts
  intermediate 10--30 keV flux.} 
\label{fig:10-30f}
\end{figure}

However, $\chi^2$ criterion alone cannot uniquely distinguish between
the three kinds of models as there are many {\em model} uncertainties
that can contribute to $\chi^2$, e.g. in both the reflection and absorption
scenarios we expect an (unknown!) range of ionization states to be
present. Hence, direct comparison of $\chi^2$ can be misleading when
the models are known to be incomplete. Both reflection and absorption
models give adequate fits to the data, and rather subtle changes in
goodness of fit are not a reliable guide as to which model best
describes the soft excess.  

A potential possibility to observationally differentiate between the
reflection and absorption models may be measuring the high energy flux
with future missions. Based on the interpolation of the best fit
models we found that the absorption model predicts the 10--30 keV flux
at the level of only 50--60\% that of the reflection dominated
model. Figure~\ref{fig:10-30f} shows the predicted model continua
(i.e., continua not affected by the warm absorption) in the considered
energy band. Nonetheless, we caution that in the most extreme cases,
such as that of 1H~0707-495, the inferred wind becomes optically thick
to electron scattering (see also Done et al. 2006). The spectrum
should then be very messy, with continuum reflection from the wind, as
well as absorption and emission from atomic features. This may wash
out the expected differences at high energies. Objects with large,
rather than extreme, soft excesses (such as PG~1211+143 where the wind
is still optically thin) may be better to test its origin.

\section{Discussion and conclusions}

We tried to determine the origin of the soft excess seen in AGN by
fitting 0.3--10~keV {\it XMM-Newton} spectra of two representative objects
with all currently proposed models, namely a separate continuum,
reflection and absorption. All these give comparably good fits to the
overall continuum given the model uncertainties (such as range of
ionisation states). Here we review the physical plausibility of
each model to help distinguish between them.

A separate soft excess component involves an emission
from an unknown physical process, with unknown 'fine-tuning' mechanism
fixing its typical energy. Moreover, this model cannot simultaneously
explain the 7~keV feature, nor is it easy to imagine how this can
produce the spectral dependence of the variability, where the {\it
rms} spectra typically peak in 0.7--3~keV band (though it is
possible to arrange: Page et al. 2004b). By contrast, both reflection and
absorption scenarios are much more plausible, as they give a physical
reason for the fixed energy of the soft excess, can reproduce the
structure around iron K and give apparently hard spectra in the
2--10~keV band, while the intrinsic spectrum remains soft. 

The reflection models can successfully fit the spectra, but the
strongest soft excesses require that the intrinsic continuum is
suppressed, while the large velocity shear to smear out the atomic
features implies extreme spin and/or emissivity (e.g. Crummy et
al. 2006). 

The alternative absorption model can reproduce the range of observed
soft excess strengths by simply changing the column density ($\sim
10^{22-23}$~cm$^{-2}$), without any requirement for the intrinsic
source to be suppressed. Thus, only a quantitative change in column is
required, rather than a qualitative change in geometry, as in the case
of reflection.  Similarly to reflection, a large velocity shear is 
required in absorption model, but since this is now identified
with the wind rather than the disc, it does not directly constrain the
space-time or energy extraction process. In pure absorption
form the model does not well fit the most extreme sharp iron features
seen at ${\sim}7$~keV.  However, emission and scattering as well as
absorption in a wind can produce a P Cygni profile in He-- and H--like
iron K$\alpha$ resonance lines. Including this gives an excellent fit
to the feature at 7~keV.  

Both reflection and absorption require that the ionisation state is
'fine-tuned' such that Oxygen is partially ionized. There is no
readily apparent reason for this in the reflection models. In fact,
the ionisation instability may actually give a physical argument
against such partially ionized species dominating reflected spectra,
making it difficult for a disc in hydrostatic equilibrium to produce
any soft excess in reflection (Done \& Nayakshin, in preparation).
By contrast, this same ionisation instability may rather naturally
produce the required range of ionisation parameter for the absorption
model (Chevallier et al. 2006). In addition, on physical
grounds we expect strong winds from high $L/L_{\rm Edd}$ accretion discs,
especially in AGN featuring a disc flux that peaks in the UV region, where
there can be strong line driving (e.g. Proga, Stone \& Kallman 2000). The
resulting, messy environment appears rather more physically plausible
than a clean reflecting disc at high $L/L_{\rm Edd}$,  and there
is growing evidence for (less extreme) relativistic outflows 
in the UV spectra of NLS1's (Leighly 2004, Green 2006). Similar material
may also be present in GBH binaries at high $L/L_{Edd}$. This might
contribute to spectral complexity at energies around iron line, but it
is unlikely to produce a soft excess at similar temperatures as in AGN
as Oxygen is probably always completely ionized due to the much higher
accretion disc temperature. 

Thus, an absorption origin seems favoured in terms of physical
plausibility, though we note that reflection from the disc should also
contribute to the spectrum at some level
(Chevallier et al. 2006). Higher energy observations soon to be
available from {\em Suzaku} may provide a more sensitive test for 
objects with the soft excess/iron feature not so large as to
require an optically thick wind.

\section*{Acknowledgements}

This work is based on observations obtained with {\it XMM-Newton}, an
ESA science mission with instruments and contributions directly funded
by ESA Member States and the USA (NASA). This research has made
extensive use of NASA's Astrophysics Data System Abstract Service. CD
acknowledges financial support through PPARC Senior fellowship. MS and
CD thank Marek Gierli\'nski and Nick Schurch for useful discussions.


\label{lastpage}

\end{document}